# Low-energy photoelectron transmission through aerosol overlayers

*Stavros Amanatidis, Bruce L. Yoder, and Ruth Signorell**

Department of Chemistry and Applied Biosciences, Laboratory of Physical Chemistry, ETH Zürich, Vladimir-Prelog-Weg 2, CH-8093 Zürich, Switzerland

*Correspondence to: rsignorell@ethz.ch.

**ABSTRACT**

The transmission of low-energy (<1.8eV) photoelectrons through the shell of core-shell aerosol particles is studied for liquid squalane, squalene, and DEHS shells. The photoelectrons are exclusively formed in the core of the particles by two-photon ionization. The total photoelectron yield recorded as a function of shell thickness (1-80nm) shows a bi-exponential attenuation. For all substances, the damping parameter for shell thicknesses below 15nm lies between 8 and 9nm, and is tentatively assigned to the electron attenuation length at electron kinetic energies of ~0.5-1eV. The significantly larger damping parameters for thick shells (> 20nm) are presumably a consequence of distorted core-shell structures. A first comparison of aerosol and traditional thin film overlayer methods is provided.

# 1. Introduction

The transport of low-energy ($\lesssim$ 10 eV) electrons through thin dielectric films has been studied for almost a century, with techniques such as low-energy photoelectron transmission (LEPET) or low-energy electron transmission (LEET) spectroscopy [1-18]. In "substrate-overlayer" studies, the dielectric is deposited on a suitable substrate under controlled conditions (in situ in vacuo, temperature controlled) to produce thin films with well-defined properties (thickness, morphology). In LEPET experiments, photoelectrons are produced in the substrate by photoionization and injected into the overlayer. The energy and angular distribution of the photoelectrons that are emitted from the overlayer into vacuum are then recorded. In LEET experiments, low-energy electrons with well-defined energy and momentum are injected into the thin film and the electron current through the film is measured. Information on various electron transport parameters, such as the electron attenuation length (EAL), the electron mean free path, scattering lengths, the escape depth, and angular scattering properties, can be retrieved from these studies. Various aspects regarding the influence of the thickness, morphology, phase, and quality of the films or cooperative and quantum effects have been addressed [1-4, 6-19]. These investigations have clearly revealed the generally complex and substance-specific scattering behaviour of low-energy electrons. Surprisingly, the concept of a "universal" EAL curve [20, 21] for dielectrics in this energy range is sometimes still being discussed even though the results from these studies clearly refute the concept for low kinetic energy electrons.

Recently, liquid microjets, aerosol particles, and molecular clusters were proposed as alternative samples for the investigation of electron transport in dielectrics [22-38]. The depth of the analysis ranges from phenomenological descriptions, to the extraction of averaged scattering properties, such as EALs, to the retrieval of detailed scattering parameters (cross sections, energetics and angular–dependences [26, 30]). In particular photoemission studies on liquid microjets and aerosol droplets finally enabled the experimental investigation of electron transport properties for volatile liquids such as water [22-26, 30]. This was not possible with the traditional overlayer method since it requires high vacuum conditions, which are incompatible with compounds with high vapor pressure. For volatiles, these new approaches have doubtless opened up new avenues. Compared with thin films, however, they come with the caveat of generally less well defined sample properties.

The present work provides a first step towards a comparison of the traditional substrate-overlayer method using flat films with an approach we refer to as the "aerosol overlayer method". Core-shell aerosol particles are used as the sample, where the core is the substrate and the shell is the overlayer. Electrons are produced in the core by photoionization and injected into the shell. The total yield of photoelectrons ejected into vacuum is then recorded as a function of the shell thickness $d$ together with the photoelectron kinetic energy (eKE), and angular distribution (PAD). We focus on the damping of the total electron yield of low-kinetic energy electrons (eKE < 2eV) through organic overlayers because a series of similar studies are available for flat thin films. Following these previous investigations, we assume exponential damping of the yield with increasing shell thickness $d$:

$$Y(d) = Y_0(d) \exp(\frac{-d}{L}) \qquad \text{Eq. (1)}$$

$Y(d)$ is the total electron yield of a core-shell particle with shell thickness $d$. $Y_0(d)$ is the total yield of photoelectrons injected from the core into the shell, which generally depends on $d$ as discussed below. $L$ is the exponential damping parameter and is often interpreted as the average electron attenuation length EAL. We have also recorded velocity map photoelectron images (VMI), which provide information on eKEs and PADs [39, 40]. However, as we have demonstrated in our previous VMI aerosol work [30, 31] a reliable analysis of these data is only possible in combination with a detailed scattering model. Corresponding thin film data that would allow a direct comparison with aerosol overlayer results are unfortunately not available. A main goal of the present study is to determine whether or not general trends found in the damping behaviour of electron yields in flat, thin layers are recovered in aerosol overlayers. We would also like to mention that aerosol overlayers have previously been used to investigate electron impact charging properties of particles and to probe organic shells on aerosol particles by secondary electron yield studies [28, 29].

## 2. Experimental setup and modelling of aerosol optical properties

### 2.1 Experimental setup

A sketch of the experimental setup is shown Fig. 1. Aerosol samples were generated by atomizing Na-benzoate (NaB; >99.5%, Sigma-Aldrich) solutions (10-20 mM) in $H_2O$ with a constant output atomizer (model TSI 3076). A fraction of the atomizer output flow (1.5

L/min) was collected and passed through a diffusion dryer (Topas DDU 570L) to remove excess humidity from the aerosol. Relative humidity levels were continuously monitored (RH sensor, Fig. 1) at the dryer outlet, and were maintained in the 0-5% range. The polydisperse aerosol then passed an electrostatic classifier (SMPS, TSI 3080) to select samples with narrowly defined mobility diameters between ~70 nm and 150 nm. The fraction of larger, doubly-charged particles exiting the classifier was minimized by selecting sizes from the large diameter edge of the size distribution. The number fraction of the doubly charged particles was estimated to be typically 15-20%. Size-selected, dry aerosol particles were then introduced into a home-built "aerosol coating device" based on the Sinclair-La Mer aerosol generator [41], which served to grow layers of low vapor pressure coating substances onto the NaB cores. Squalane ($C_{30}H_{62}$, >99%, Fluka), squalene ($C_{30}H_{50}$; >99%, Acros), and di-ethyl-hexyl-sebacate (DEHS; $C_{26}H_{50}O_4$, >97%, Aldrich) were used as coating substances. In the coating device, the incoming flow of core particles was gradually heated to ~250°C and mixed with a coating vapor-laden $N_2$ flow, generated by bubbling a precise $N_2$ mass flow (mass flow controller, MFC, Alicat Scientific) through a temperature controlled container with liquid coating material. Upon cooling in a laminar flow, super-saturated vapors of the coating substance condense onto the Na-benzoate particles which serve as nuclei for heterogeneous condensation. The particle size can be adjusted by regulating the temperature of the coating substance and the mass flow of the coating vapor introduced into the coating device. Particle sizes were probed with a second Scanning Mobility Particle Sizer (SMPS, TSI 3938), from which the core (~38 – 75nm) and shell thicknesses (~1 – 80nm) could be deduced. A Condensation Particle Counter (CPC, TSI 3775) positioned downstream of an additional dilution stage was employed to continuously monitor the particle number concentration at the outlet of the coating device.

A portion of the aerosol output from the coating device was introduced into vacuum by means of a home-built aerodynamic lens (ADL) system [42, 43]. The resulting continuous aerosol beam traversed a differential pumping chamber before entering the detection chamber which was held at ≤1x10$^{-6}$ mbar during the experiment. The imaging setup used for this study is similar to the ones we have previously described in detail [30, 31, 44]. The NaB cores were photoionized by two-photon absorption of 266nm (4.66eV) laser radiation from ~8ns pulses (Quantel Ultra) at a repetition rate of 20Hz. All coating materials are transparent at this wavelength (see refractive index data below) and were thus not photoionized. The threshold ionization energies $IE_{th}$ of NaB, squalane, squalene, and DEHS are ~7.5eV (this work);

~8.4eV [45] ; ~6.9eV [45], and ; ~8.4eV [45], respectively. We checked that droplets of the pure coating substances, which are transparent to 266nm light, do not produce any detectable 2-photon ionization signal. The ionizing light was linearly polarized parallel to the detection plane, as indicated in Fig. 1 by the double headed arrow at the point of intersection between the aerosol beam and the laser beam (direction of propagation perpendicular to the plane of the figure). Inside a pair of concentric mu-metal cylinders, the generated photoelectrons were extracted vertically with respect to the aerosol beam by a three plate extractor and collected with a position sensitive Electron detector (Photonis APD 2 40/12/10/8 I 60:1 MGO 6'' FM P43).

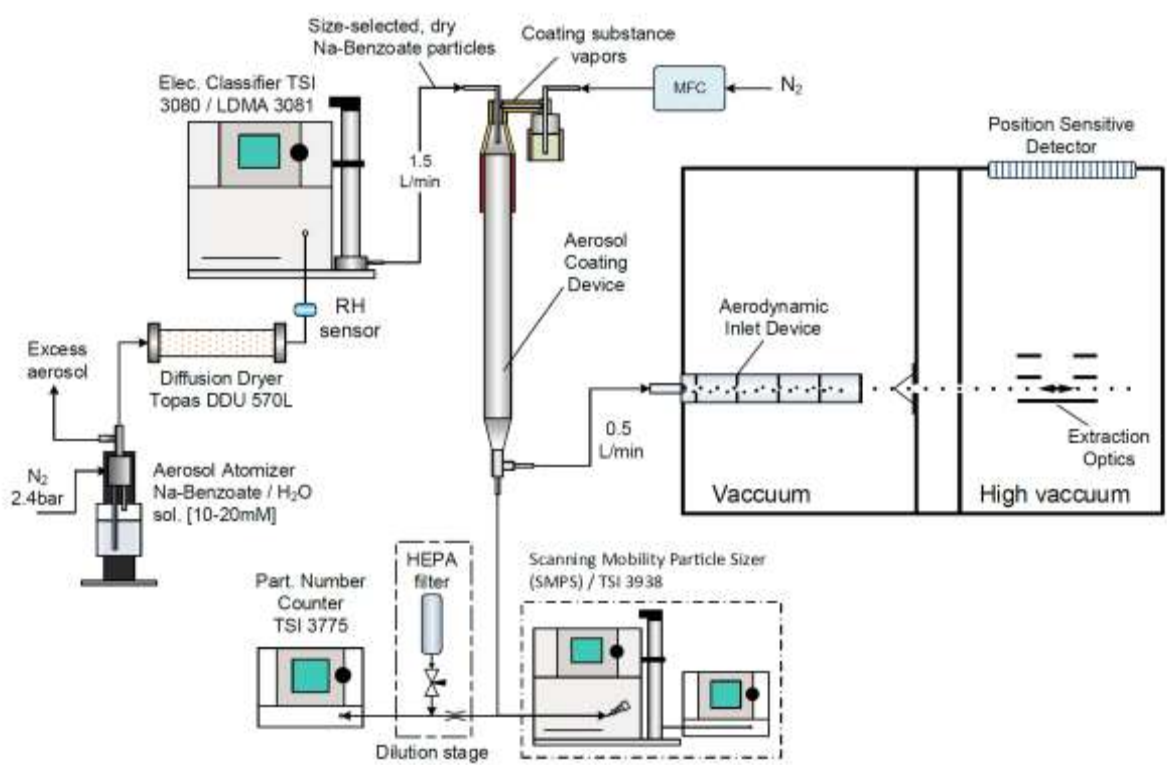

**Figure 1:** Scheme of the experimental setup consisting of devices for aerosol formation (aerosol atomizer, aerosol coating device), devices for aerosol selection and characterization (SMPS), the VMI photoelectron spectrometer with an ADL for the transfer of aerosols into the spectrometer and a laser for photoionization of the aerosol particles.

For the study of photoelectron yields as a function of shell thicknesses, spatial map imaging (SMI) and velocity map imaging (VMI) [39, 40] methods of photoelectron detection were compared. VMI detection produced an intense signal at the image center spot (near threshold photoelectrons) that could make the method prone to saturation effects. SMI detection spreads the photoelectron signal over a larger area of the detector than VMI. Nevertheless, for the signal levels used in this work, no saturation effects were observed for either method. We used SMI detection for the yield measurement as it produced the highest signal to background levels and was the least prone to saturation effects. All yield measurements are background corrected. We also recorded VMIs to retrieve semi-quantitative information on the electron kinetic energy (eKE) distributions. Correct, quantitative eKE distributions can only be extracted from reconstructed VMIs for cylindrically symmetric arrangements [46, 47]. For aerosol particles, this symmetry is broken by shadowing and nanofocussing effects (see refs. [30, 31] for details). The usual reconstruction methods are strictly no longer applicable and can only provide semi-quantitative information on the true eKE distributions. An estimate of the falsification of the true eKE distributions upon reconstruction is provided in section 4. Correct, quantitative eKE information can be extracted from aerosol VMIs only by means of detailed optical and electron scattering models [30].

## 2.2 Modeling of aerosol optical properties

The interaction of aerosol particles with the ionizing radiation is strongly influenced by the finite size of the particles (see shadowing and nanofocusing in ref. [30] and ref. [48]). These optical confinement effects can result in a pronounced increase (nanofocusing) of the volume averaged light intensity inside the particle compared with flat, thin films and bulk samples. This is the case in our experiments. The effects strongly depend on the particle size, architecture, and optical properties. Since the photoionization probability in the NaB core is proportional to the square (two-photon process) of the local light intensity $I$ in the core, the yield of the electrons injected from the core into the shell is not constant for a fixed core size, but varies with the shell thickness. These variations introduce substantial corrections to the yield of the bare core, $Y_0(d)$, which must be taken into account for correct relative yields $\frac{Y(d)}{Y_0(d)}$ (Fig. 2a). We have calculated the volume averaged light intensity in the core for all core radii and shell thicknesses by solving the Maxwell equations and used these data to properly scale $Y_0$:

$$Y_0(d) = Y_0(0) \frac{\langle I^2(d) \rangle}{\langle I^2(0) \rangle} \qquad \text{Eq. (2)}$$

The calculations, which were performed with the ADDA package [49], require refractive index data for both core and shell. The real ($n$) and imaginary parts ($k$) of the complex refractive index of NaB, squalane, squalene, and DEHS at 266 nm were derived by Kramers-Kronig inversion [50] from UV/VIS spectra we recorded for all four substances. This resulted in the following values of the complex index of refraction $n+i \cdot k$: NaB: $1.70+i \cdot 0.03$; squalane: $1.55+i \cdot 10^{-5}$; squalene: $1.61+i \cdot 10^{-4}$;, and DEHS: $1.63+i \cdot 10^{-5}$. The values for squalane, squalene, and DEHS agree with previous literature values [51, 52].

## 3. Results

The full squares in Fig. 2a show the experimental relative electron yield $\frac{Y(d)}{Y_0}$ as a function of squalene shells with thicknesses $1 \lesssim d \lesssim 80$nm for seven different NaB core radii $r$ between ~38 and 75nm and assuming a constant core yield, i.e. $Y_0 = Y_0(0)$. Similar results were obtained for squalene and DEHS coatings (Table 1). Obviously, the data for similar $d$ scatter pronouncedly. A closer inspection of the data points reveals a correlation between the relative electron yield and the substrate core size $r$; namely that $\frac{Y(d)}{Y_0}$ is smaller for larger $r$. This correlation is expected for small aerosol particles because the average light intensity in the core where the electrons are formed sensitively depends on the overall particle size ($r+d$), its architecture ($\frac{d}{r}$), and the optical properties of core and coating [30, 31, 48]. In contrast to flat substrate-overlayers, where such optical finite-size effects do not occur, the yields $Y_0$ of aerosols particles need to be corrected for these optical effects. To account for this, we solved Maxwell's equations for all aerosol particles (section 2.2 and next paragraph) to determine the correct core yield according to Eq.(2). The corrected relative yields are shown as open circles in Fig. 2a. The scatter in the data is significantly reduced and correlations with the core size are removed in the corrected data. The remaining scatter (0.01) in the data serves as a measure of their reproducibility. It mainly arises from background variations, the limited accuracy of the particle radii (size-selection), and uncertainties in their number density (transmission through the ADL system into vacuum). The comparison of the uncorrected and corrected data in Fig. 2a reveals that the data for thick overlayers could only be recorded because the overall photoelectron yield increased to viable signal levels as a consequence of the he light enhancement in the core.

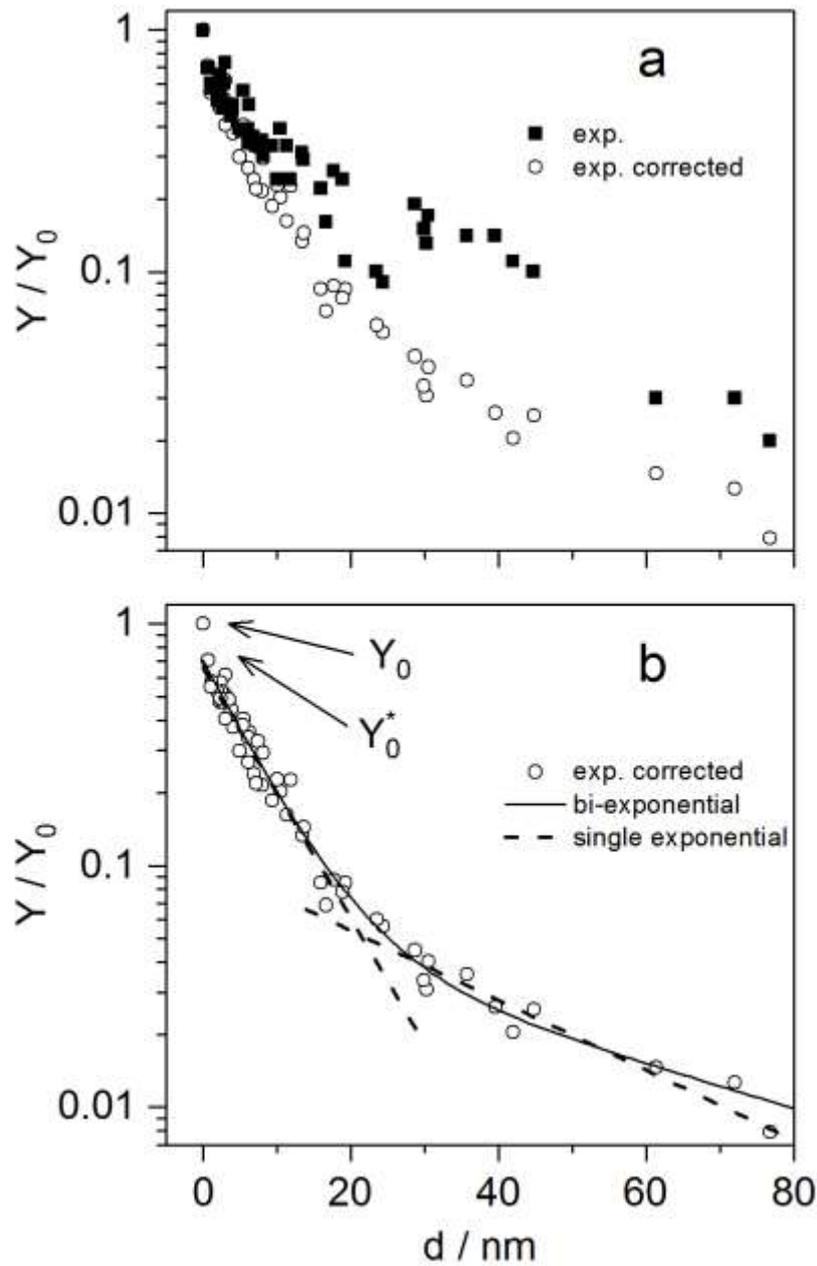

**Figure 2:** Semi-log plot of the relative electron yield $\frac{Y}{Y_0}$ as a function of the shell thickness $d$ for the example of squalane coatings. a) Full squares: original experimental data assuming constant core yield $Y_0=Y_0(0)$. Open circles: experimental data after correction for optical effects in aerosol particles, (Eq. (2)). b) Corrected experimental data (open circles); bi-exponential fit (full line); fit with two single exponentials (dashed lines).

We observed two general trends in the influence of optical confinement effects. The first is the above mentioned core size dependence for a given shell thickness. The corrections are typically larger (up to a factor of ~2-3) for smaller NaB cores because particles with larger $\frac{d}{r}$ ratios couple light more efficiently into the core compared with the bare (uncoated) core. The second trend arises from a general increase in the average light intensity in the NaB core with increasing coating thickness. As a consequence, more electrons (higher core yield $Y_0$) are formed in a core with a thicker coating than in the same bare core alone; corrections thus tend to be larger for thick coatings than for thin ones (up to a factor of ~2-3). For example, a coating of thickness $d \sim 3$nm on a core of $r = 74$nm hardly changes the electron yield $Y_0$ of the core compared with the bare core so that essentially no correction is required. A 77nm shell, by contrast, increases $Y_0$ of the coated core compared with the bare core by a factor of ~2.5. The latter case is illustrated in Fig. 3, which shows the square of the light intensity (two-photon ionization) in a plane through the center of the particle spanned by the directions of propagation and polarization of the light. The upper panel shows the bare core and the lower panel the coated particle. The example visualizes the situation where the average light intensity in the core is increased in the coated particle mainly because of nanofocusing by the shell, which produces a hot spot of the light intensity in the right half of the particle. The systems studied here exhibit a general propensity for overall larger particles ($r+d$) to collect more light than smaller particles (see also discussion on nanofocusing in ref. [30].). Note that such general trends depend on the specific optical properties and the sizes of core and shell so that a simple generalization for other particles is not possible.

Fig. 2b shows the experimental relative yield $\frac{Y(d)}{Y_0(d)}$ corrected for optical effects according to (Eq. (2)) and compares it with two different fits assuming exponential damping by the shell (Eq. (1)). The non-linear behavior of the experimental data in the semi-log plot shows that a single exponential function is not sufficient to describe the damping behavior over the whole range of shell thicknesses. Two exponential functions are required to describe the data. We use two models, which we refer to as model 1 and model 2, respectively. Model 1 uses two single exponentials to describe the average damping in the regions $d < 15$ nm and $d > 20$ nm, respectively (dashed lines in Fig. 2b). The corresponding damping lengths are referred to as $L_{thin}$ and $L_{thick}$. Model M2 is a bi-exponential fit with damping constants $L_{thin}$ and $L_{thick}$. The fit results for the different coating substances squalane, squalene, and DEHS are summarized in Table 1. For model M2, we have performed a detailed sensitivity analysis for $L_{thin}$ and $L_{thick}$, which accounts for uncertainties in the recorded electron yields, the particle sizes, the refractive index data used to correct for optical effects, and the influence of doubly-charged particles generated in the SMPS (section 2.1). The resulting limiting values for $L_{thin}$ and $L_{thick}$ are indicated in Table 1 as subscripts and superscripts, respectively. We note that this analysis does not account for sources of systematic errors, such as non-spherical particles and incomplete coatings, which are currently impossible to quantify.

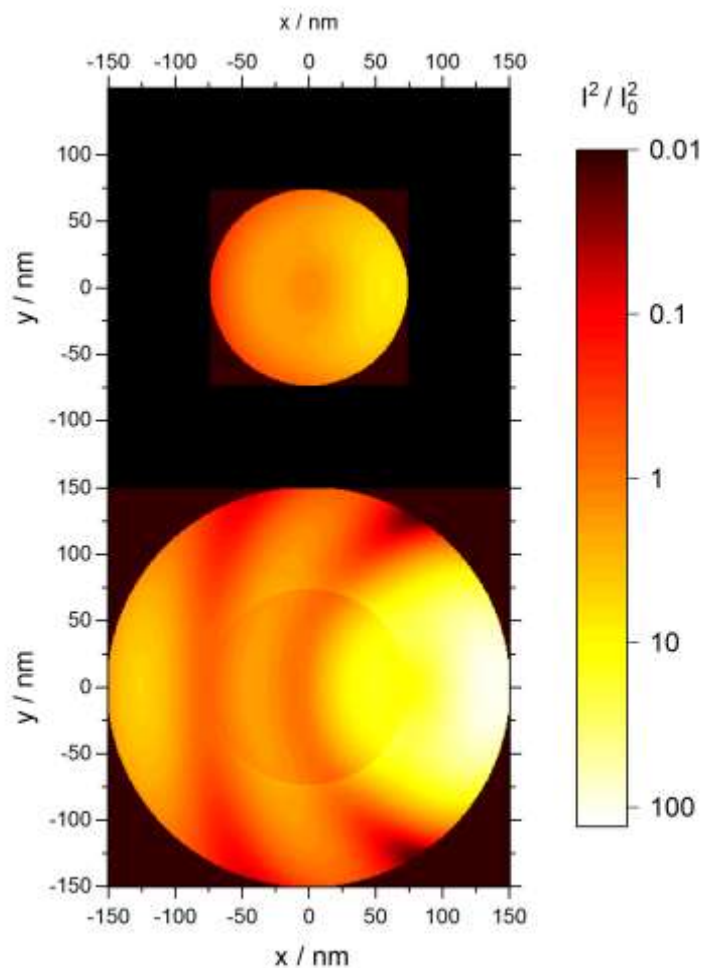

**Figure 3:** Square of the local light intensity $I^2$ relative to the square of the light intensity $I_0^2$ of the incoming light (not shown) for 266nm light in a plane through the center of the particle that is spanned by the directions of polarization (*y*) and propagation (*x*) of the light. Note the logarithmic colour scale. Upper panel: bare NaB core with *r* = 74nm. Lower panel: NaB core with *r* = 74nm coated with a squalane shell of *d* = 77nm.

## 4. Discussion

The general trends observed in Fig. 2b for aerosol overlayers agree well with those previously found for total photoyield measurements on flat thin films [1, 4, 13-16, 53, 54]. The yield $Y_0$ measured for a bare core is higher than the corresponding yield $Y_0^*$ extrapolated by the exponential damping models. The same behavior was observed for thin flat films, and was explained by scattering at the substrate-film interface [4, 15]. The presence of the shell leads to a reduction of the electron yield emitted by the coated core ($Y_0^*$) compared with the yield emitted by the bare core ($Y_0$) because of partial back-reflection of electrons into the core at the core-shell interface. In agreement with flat thin films [15], we also find that two exponentials are required to reproduce the experimental data over the whole range of coating thicknesses; one that describes the damping for thin ($L_{thin}$) and another one that describes the damping for thick coatings ($L_{thick}$). This was explained to arise from the more efficient elimination of "higher" energetic electrons (eKEs in the range of a few eV) in the first few monolayers of the film compared with the "lower" energetic electrons (eKEs on the order of ~0.4eV), which experience substantial damping only in thick films. $L_{thin}$ and $L_{thick}$ were thus interpreted to represent the EALs for higher and lower energetic electrons, respectively. The energy-dependent damping was rationalized as follows [5, 15, 55, 56]. Electrons that are emitted from the substrate into the overlayer are distributed over a range of initial eKEs covering both lower and higher eKEs. The more efficient losses in the thin layers were explained by energy losses of the higher energetic electrons of the initial eKE distribution to high frequency *intramolecular* vibrations, and thus $L_{thin}$ was interpreted as the EAL of higher kinetic energy electrons. These intramolecular scattering processes are comparatively efficient with an average loss per scattering event on the order of typical intramolecular vibrational energies (e.g. CH-stretching, vibrational quanta ~0.35eV). As a result, the higher energetic electrons reach energies below typical intramolecular vibrational energies after passing only a few monolayers, where they have been converted to lower energetic electrons. $L_{thick}$ was interpreted to correspond to the EAL of such lower energetic electrons with energies below typical intramolecular vibrations. The dominant scattering mechanisms for these electrons occurs through *intermolecular* vibrations and elastic processes. Typical intermolecular vibrational energies are on the order of 0.01eV. Many phonon collisions are thus required to reduce the kinetic energy of these electrons to thermal energies (0.025 eV at room temperature). This explanation is qualitatively consistent with larger values of $L_{thick}$ compared with $L_{thin}$.

The qualitative agreement of our aerosol data with the results from flat thin films suggests a similar explanation for the damping behavior of aerosol overlayers; namely that the values of $L_{thin}$ and $L_{thick}$ extracted for model M1 in Table 1 represent the EALs at "higher" and "lower" eKEs, respectively. In addition, the subscripts in Table 1 obtained from the bi-exponential fit of model M2 provide firm estimates of the lower bound for $L_{thick}$ and thus of the corresponding EAL. These lower bounds are consistent with the average values $L_{thick}$ of model M1. Model M2 also reveals a strong correlation between $L_{thin}$ and $L_{thick}$ so that $L_{thin}$ does not directly compare to the thin coating parameter of model M1. While model M2 does approach model M1 in the limit of thick coatings, model M2 describes a gradual change of the effective damping length in the range of thin coatings. Only the average value would correlate with $L_{thin}$ of model M1. A rough estimate of what "higher" and "lower" electron kinetic energies mean in the aerosol case can be obtained from an eKE spectrum of bare NaB cores. Fig. 4 shows the eKE spectrum of bare cores retrieved from a reconstructed VMI image (thick black line). Even though it is not possible to determine accurate eKE distributions from such reconstructed VMIs of aerosol particles (section 2.1 and discussion below) they still provide useful semi-quantitative information on the eKE distribution. Fig. 4 shows that the NaB cores emit electrons over an eKE range from ~0 to 1.8eV. The majority of the emitted electrons, however, have eKEs between ~ 0.2 and 1eV and the number of electrons with eKEs > 1.5eV is negligible. From the general scattering behavior of electrons with sub-electronic kinetic energies (previous paragraph) one would thus tentatively assign the term "higher" eKEs to the range around ~0.5-1eV, while "lower" eKEs would approximately cover eKEs $\lesssim$ 0.4eV. To test whether thicker layers indeed lead to a depression of the fraction of higher eKEs we have recorded eKE spectra for different layer thicknesses. Fig. 4 compares the eKE distribution recorded for a bare core with that resulting from the same core coated with a 20 nm shell. We find almost unchanged distributions; i. e. a damping of the electron signal that hardly depends on the eKE. In fact, the coated particle shows a distribution slightly shifted to higher eKE values. However, the latter should be not overinterpreted in view of the reconstruction issues discussed above (section 2.1). As discussed in the next paragraph, this result questions the interpretation of $L_{thick}$ as a low-eKE EAL.

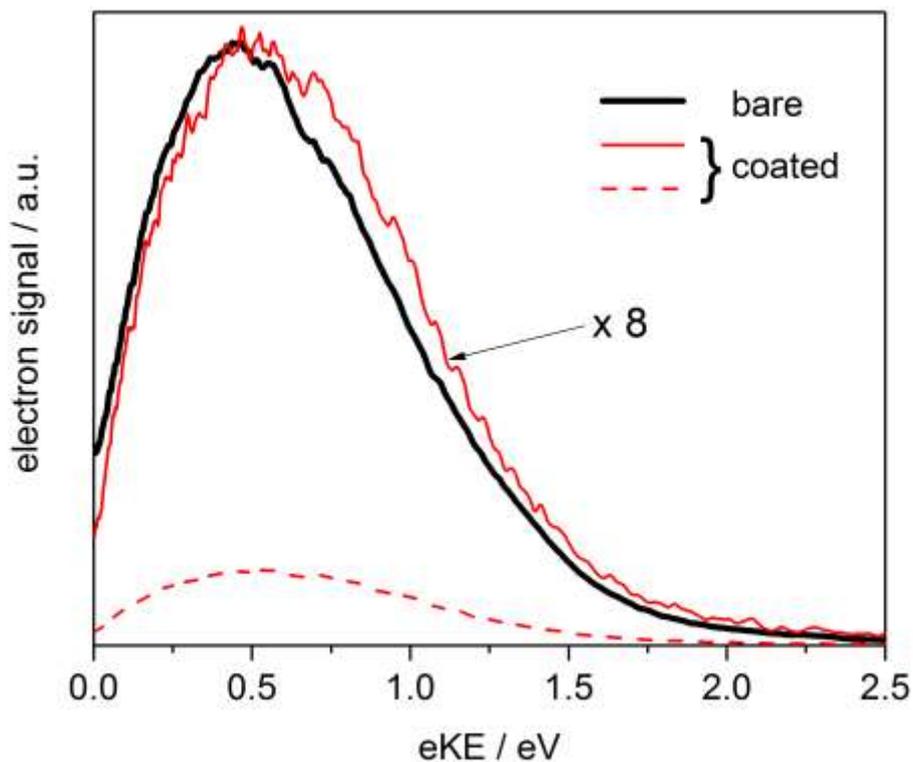

**Figure 4:** eKE distributions of bare NaB cores (thick black line) and of NaB particles with a 20 nm coating of squalane (thin red lines). The dashed red line is the unscaled spectrum and the full red line is the scaled spectrum.

The interpretation of $L_{thin}$ as an average EAL at eKEs in the range of ~0.5-1eV still seems reasonable. It is also plausible that almost identical values result for the three different compounds ($L_{thin}$ in Table 1) because of the similar structures with high fractions of CH moieties, which presumably dominate scattering in this eKE range. Even though a direct comparison with other room temperature data for the present coating substances is not possible, $L_{thin}$ is of similar magnitude as corresponding values for solid hydrocarbon films [15]. The interpretation of $L_{thick}$ as the EAL at very low eKEs is definitely questionable. From Fig. 4 it is clear that the slower damping by thick shells does not simply reflect an increase of the effective EAL at lower eKEs. Furthermore, compared with corresponding values for other solid hydrocarbon layers, the $L_{thick}$ values determined in the present work seem rather large (note, however, that they are not very well determined). Here, we just mention two potential phenomena that could falsify the true EALs and lead to artificially large values. Since our coatings are liquid, it is not clear whether the core is really positioned in the center of the particle. This is particularly problematic for very thick shells *(d ~ r)*. For pronouncedly off-centered core positions one would expect to measure higher electron yields compared with nicely centered cores, resulting in too high EAL values for the assumed coating thickness. Similar effects could arise when the barrier the electrons have to overcome to escape into vacuum is smaller than or similar to the thermal energy (~0.025 eV). Thermal electrons, which can not only lose but also gain kinetic energy in scattering events, have on average constant kinetic energy and are transported by diffusion. For low escape barriers (i.e. low or negative electron affinities), they can escape by diffusion and produce a small photoelectron "background" that would result in artificially high EALs. The escape barriers for the particles studied here are not known, but given the negative electron affinities of some hydrocarbons [4], the diffusive escape of thermal electrons might contribute to the photoelectron signal measured for very thick coatings. For the 20 nm shell shown in Fig.4 this contribution seems to be less significant, since the measured eKE distributions do not appear to be distorted by thermal electrons (near eKE=0). In principle, we cannot exclude a complicated interplay of various energy-dependent scattering contributions as an explanation for the observed double-exponential behavior. But, at present we consider imperfections of the core-shell architecture as a plausible explanation for the less effective damping by thick coatings.

Fig. 4 might raise the question why we just rely on the total photoelectron yield. One might be tempted to retrieve the energy-dependence of the EAL from the damping of the electron signal recorded at different eKE values (Fig. 4). As has been discussed extensively in the literature, however, such an approach would be fundamentally flawed as it neglects the cascade of energy transfer down the eKE scale by scattering [4, 54, 57]. The exception is the topmost eKE interval from which electrons can only be lost. Studies on flat thin films have shown that accurate data can be extracted from eKEs distributions using this "topmost energy interval" method [4, 54, 57]. Applying it to the present case is, however, problematic. The reason lies in the velocity map imaging method we used for this study. As explained in section 2.1, reconstructed VMIs of aerosol particles do not allow one to retrieve true and thus accurate eKE distributions because of symmetry issues. As a result, reconstructed images are distorted compared with the true distributions and do thus not provide accurate electron yields as a function of eKE. We have performed simulations on model aerosol particles using a detailed scattering model [26, 30] to quantify these effects. These simulations reveal that the damping of the electron signal as a function of eKE derived from reconstructed images can systematically deviate by a factor of up to 2 from the true values. Consequently, the EALs derived by the topmost energy interval method would be correspondingly uncertain. For this reason, we only use the total yield information, which is not biased by reconstruction. However, we would also like to add here that the information contained in the raw VMIs, i. e. not reconstructed, is in principle very detailed and useful. But, it can only be extracted by means of a sophisticated scattering models that accounts for all different scattering processes, their cross-sections, energetics, and angular-dependences. We have successfully demonstrated such an approach for VMIs of liquid water droplets [30]. However, this would be a very demanding task which goes far beyond the present goal of a first evaluation of the aerosol overlayer method.

Even though a direct comparison of our aerosol approach with traditional substrate-overlayer studies for the same compounds is not possible, we can already make a first general comparison of the two methods at this point. The electron yields of our aerosol approach (Fig. 2b) scatter more pronouncedly than typical yields extracted from flat overlayers [4, 15, 57], which hints at a poorer reproducibility of the aerosol approach. The way how the coated aerosol particles are formed (atomizer, aerosol coating device), size-selected, and characterized (ex-situ with an SMPS), and the way how they are transferred into vacuum (ADL) are very likely to result in less reproducible samples compared with thin layers, where comparatively high control and thus high reproducibility is achievable. The limited control over the sample properties also results in less accurate values of parameters such as EALs. Not only the determination of the exact particle number concentration and the exact size of core and shell can be an issue here, but also the missing information on the particles' shape and architecture. For example, it is by no means clear whether the particles are indeed spherical as assumed here or whether the core is symmetrically surrounded/enclosed by the shell. In addition, aerosol particles have a very high, size-dependent curvature – an issue that does not arise for flat, thin films. This could lead to structural modifications of the coating and thus alter the EAL. Similarly, the generation methods used for aerosols are more prone to impurities than the generation methods for thin film (in-situ in vacuo). Impurities might also alter the damping properties. Finally, photoexcitation needs careful corrections of the complicated optical effects that arise from the finite size of particles. We have demonstrated that such corrections are possible, but their quality relies on accurate information about size, shape, and refractive indices.

## 5. Summary

We have studied the attenuation of low-kinetic energy photoelectrons through thin aerosol coatings for the three liquid hydrocarbons squalane, squalene, and DEHS. The liquids form overlayers on solid Na-benzoate core particles, which act as the photoelectron source. The ionization of the cores by two-photon ionization at 4.6eV photon energy ensures that no electrons are formed in the overlayers, which are transparent at this photon energy. The finite-size of aerosol particles results in a strong enhancement of the light intensity and thus the electron yield in the core. This signal enhancement allows us to investigate also very thick layers (several ten nanometers). An average electron attenuation length of 8-9nm at electron kinetic energy of ~0.5-1eV is determined from the exponential damping in thin layers (< 15nm). The less efficient damping in thick layers seems to be the result of distorted core-shell structures.

On the one hand, the aerosol overlayer approach seems less accurate and reproducible compared with the traditional thin, flat film overlayer method, mainly a consequence of the more limited control over sample preparation and the more difficult characterization of aerosol samples. On the other hand, the aerosol approach is comparatively simple. Furthermore, it could principally provide very detailed information about the underlying scattering processes when it is combined with angle-resolved photoelectron spectroscopy (velocity map imaging). However, as we have shown in previous studies on water aerosol droplets this requires the help of detailed scattering models that explicitly consider all different scattering channels, their cross sections, energetics, and angular-dependence [26, 30].

**Acknowledgment:** We thank David Stapfer and Markus Steger for technical support for the setup of the coating device, Daniel Zindel for the measurement of the UV/VIS spectra of NaB, squalane, squalene, and DEHS, and David Luckhaus for help with the calculations. Financial support was provided by the ETH Zürich and the Swiss National Science Foundation (project 200020_159205).

# Tables

Table 1: Exponential damping parameters $L_{thin}$ and $L_{thick}$ for three different coating materials. Model 1: Single exponential fits in the regions $d < 15$ nm ($L_{thin}$) and $d > 20$ nm ($L_{thick}$), respectively. The value in parentheses is one standard deviation. The parameter $L_{thin}$ is tentatively assigned to the EAL at eKEs ~0.5-1eV, while $L_{thick}$ describes the less effective damping by very thick layers, presumably as a consequence of distorted core-shell structures (see text). Model 2: Bi-exponential fit. The subscript and superscript represent the range of uncertainties (correlated lower and upper bounds). ">> x" indicates values much larger than the thickest coating. The subscripts provide an estimate of the lower bound for $L_{thick}$ (see text).

| Material | Model | $L_{thin}$ / nm | $L_{thick}$ / nm |
|---|---|---|---|
| squalane | M1 | 8.4(0.5) | 30(3) |
|  | M2 | $7.1^{9.0}_{5.0}$ | $47^{>>80}_{20}$ |
| squalene | M1 | 9.2(0.6) | 68(17) |
|  | M2 | $8.1^{10.0}_{5.4}$ | $>> 60^{>>60}_{30}$ |
| DEHS | M1 | 8.4(0.6) | 25(6) |
|  | M2 | $5.9^{9.0}_{3.8}$ | $30^{>>60}_{20}$ |

**REFERENCES**

[1] R. Naaman, and L. Sanche, Chem. Rev. **107**, 1553 (2007).

[2] A. Nitzan, Annu. Rev. Phys. Chem. **52**, 681 (2001).

[3] R. Naaman, and Z. Vager, Acc. Chem. Res. **36**, 291 (2003).

[4] J. Bernasconi, E. Cartier, and P. Pfluger, Phys. Rev. B **38**, 12567 (1988).

[5] E. Cartier, and P. Pfluger, Physica Scripta **T23**, 235 (1988).

[6] N. Ueno, and K. Sugita, Phys. Rev. B **42**, 1659 (1990).

[7] M. Rei Vilar, M. Schott, and P. Pfluger, J. Chem. Phys. **92**, 5722 (1990).

[8] T. Goulet, V. Pou, and J. P. Jay-Gerin, J. Electron. Spectrosc. Relat. Phenom. **41**, 157 (1986).

[9] J. P. Jay-Gerin, B. Plenkiewicz, P. Plenkiewicz, G. Perluzzo, and L. Sanche, Solid State Commun. **55**, 1115 (1985).

[10] R. D. Birkhoff, J. M. Heller, L. R. Painter, J. C. Ashley, and H. H. Hubbell, J. Chem. Phys. **76**, 5208 (1982).

[11] L. R. Painter, E. T. Arakawa, M. W. Williams, and J. C. Ashley, Radiat. Res. **83**, 1 (1980).

[12] J. C. Ashley, C. J. Tung, and R. H. Ritchie, IEEE Transactions on Nuclear Science **25**, 1566 (1978).

[13] S. Hino, N. Sato, and H. Inokuchi, Chem. Phys. Lett. **37**, 494 (1976).

[14] J. T. J. Huang, and J. L. Magee, J. Chem. Phys. **61**, 2736 (1974).

[15] Y. C. Chang, and W. B. Berry, J. Chem. Phys. **61**, 2727 (1974).

[16] R. A. Holroyd, B. K. Dietrich, and H. A. Schwarz, J. Phys. Chem. **76**, 3794 (1972).

[17] M. Michaud, A. Wen, and L. Sanche, Radiat. Res. **159**, 3 (2003).

[18] D. Yoshimura, H. Ishii, Y. Ouchi, E. Ito, T. Miyamae, S. Hasegawa, K. K. Okudaira, N. Ueno, and K. Seki, Phys. Rev. B **60**, 9046 (1999).

[19] A. Kadyshevitch, S. P. Ananthavel, and R. Naaman, J. Chem. Phys. **107**, 1288 (1997).

[20] C. J. Powell, J. Electron. Spectrosc. Relat. Phenom. **47**, 197 (1988).

[21] M. P. Seah, Surf. Interface Anal. **44**, 497 (2012).

[22] N. Ottosson, M. Faubel, S. E. Bradforth, P. Jungwirth, and B. Winter, J. Electron. Spectrosc. Relat. Phenom. **177**, 60 (2010).

[23] S. Thürmer, R. Seidel, M. Faubel, W. Eberhardt, J. C. Hemminger, S. E. Bradforth, and B. Winter, Phys. Rev. Lett. **111**, 173005 (2013).

[24] Y.-I. Suzuki, K. Nishizawa, N. Kurahashi, and T. Suzuki, Phys. Rev. E **90**, 010302 (2014).

[25] Y. Yamamoto, S. Karashima, S. Adachi, and T. Suzuki, J. Phys. Chem. A **120**, 1153 (2016).

[26] D. Luckhaus, Y. Yamamoto, T. Suzuki, and R. Signorell, Science Advances, in press (2017).

[27] G. Olivieri, K. M. Parry, C. J. Powell, D. J. Tobias, and M. A. Brown, J. Chem. Phys. **144**, 154704 (2016).

[28] P. J. Ziemann, and P. H. McMurry, Aerosol Sci. Technol. **28**, 77 (1998).

[29] P. J. Ziemann, P. Liu, D. B. Kittelson, and P. H. McMurry, J. Phys. Chem. **99**, 5126 (1995).

[30] R. Signorell, M. Goldmann, B. L. Yoder, A. Bodi, E. Chasovskikh, L. Lang, and D. Luckhaus, Chem. Phys. Lett. **658**, 1 (2016).

[31] M. Goldmann, J. Miguel-Sánchez, A. H. C. West, B. L. Yoder, and R. Signorell, J. Chem. Phys. **142**, 224304 (2015).

[32] M. J. Berg, K. R. Wilson, C. M. Sorensen, A. Chakrabarti, and M. Ahmed, J. Quant. Spectrosc. Radiat. Transfer **113**, 259 (2012).

[33] F. Süßmann, L. Seiffert, S. Zherebtsov, V. Mondes, J. Stierle, M. Arbeiter, J. Plenge, P. Rupp, C. Peltz, A. Kessel, S. A. Trushin, B. Ahn, D. Kim, C. Graf, E. Rühl, M. F. Kling, and T. Fennel, Nat. Commun. **6**, 7944 (2015).

[34] F. Calegari, S. Zherebtsov, L. Seiffert, Q. Liu, A. Trabattoni, P. Rupp, M. Castrovilli, G. Sansone, V. Mondes, I. Halfpap, C. Graf, E. Rühl, M. Nisoli, T. Fennel, and M. F. Kling, in *2016 Conference on Lasers and Electro-Optics (CLEO)* (OSA, 2016), pp. 1.

[35] J. L. Ellis, D. D. Hickstein, W. Xiong, F. Dollar, B. B. Palm, K. E. Keister, K. M. Dorney, C. Ding, T. Fan, M. B. Wilker, K. J. Schnitzenbaumer, G. Dukovic, J. L. Jimenez, H. C. Kapteyn, and M. M. Murnane, J. Phys. Chem. Lett. **7**, 609 (2016).

[36] S. Hartweg, B. L. Yoder, G. A. Garcia, L. Nahon, and R. Signorell, Phys. Rev. Lett., in press (2017).

[37] M. Winkler, V. Myrseth, J. Harnes, and K. J. Børve, J. Chem. Phys. **141**, 164305 (2014).

[38] C. Zhang, T. Andersson, M. Förstel, M. Mucke, T. Arion, M. Tchaplyguine, O. Björneholm, and U. Hergenhahn, J. Chem. Phys. **138**, 234306 (2013).

[39] D. W. Chandler, and P. L. Houston, J. Chem. Phys. **87**, 1445 (1987).

[40] A. Eppink, and D. H. Parker, Rev. Sci. Instrum. **68**, 3477 (1997).

[41] D. Sinclair, and V. K. La Mer, Chem. Rev. **44**, 245 (1949).

[42] P. Liu, P. J. Ziemann, D. B. Kittelson, and P. H. McMurry, Aerosol Sci. Technol. **22**, 293 (1995).

[43] P. Liu, P. J. Ziemann, D. B. Kittelson, and P. H. McMurry, Aerosol Sci. Technol. **22**, 314 (1995).

[44] B. L. Yoder, A. H. C. West, B. Schläppi, E. Chasovskikh, and R. Signorell, J. Chem. Phys. **138**, 044202 (2013).

[45] H. Koizumi, K. Lacmann, and W. F. Schmidt, J. Electron. Spectrosc. Relat. Phenom. **67**, 417 (1994).

[46] V. Dribinski, A. Ossadtchi, V. A. Mandelshtam, and H. Reisler, Rev. Sci. Instrum. **73**, 2634 (2002).

[47] B. Dick, PCCP **16**, 570 (2014).

[48] C. F. Bohren, and D. R. Huffman, *Absorption and scattering of light by small particles* (Wiley, New York, 1998), Wiley science paperback series.

[49] M. A. Yurkin, V. P. Maltsev, and A. G. Hoekstra, J. Quant. Spectrosc. Radiat. Transfer **106**, 546 (2007).

[50] J. P. Hawranek, P. Neelakantan, R. P. Young, and R. N. Jones, Spectrochim. Acta, Part A **32**, 85 (1976).

[51] L. R. Painter, J. S. J. S. Attrey, H. H. Hubbell Jr., and R. D. Birkhoff, J. Appl. Phys. **55**, 756 (1984).

[52] A. Pettersson, E. R. Lovejoy, C. A. Brock, S. S. Brown, and A. R. Ravishankara, J. Aerosol Sci. **35**, 995 (2004).

[53] E. Cartier, P. Pfluger, J.-J. Pireaux, and M. Rei Vilar, Appl. Phys. A **44**, 43 (1987).

[54] E. Cartier, and P. Pfluger, Phys. Rev. B **34**, 8822 (1986).

[55] W. F. Schmidt, and A. O. Allen, J. Phys. Chem. **72**, 3730 (1968).

[56] A. Mozumder, and J. L. Magee, J. Chem. Phys. **47**, 939 (1967).

[57] P. Pfluger, H. R. Zeller, and J. Bernasconi, Phys. Rev. Lett. **53**, 94 (1984).